\documentclass[runningheads]{llncs}
\usepackage[T1]{fontenc}
\usepackage{graphicx}
\usepackage[colorlinks=true, linkcolor=black, urlcolor=blue]{hyperref}
\usepackage{multirow}
\usepackage{amsmath, xparse}
\usepackage{orcidlink}
\usepackage{algorithm}
\usepackage{algorithmic}
\begin{document}
\title{HyperCMR: Enhanced Multi-Contrast CMR Reconstruction with Eagle Loss}
\titlerunning{Efficient CMR Reconstruction}

\author{Ruru Xu\inst{1} \and
Caner Özer\inst{1,2,3} \and
Ilkay Oksuz\inst{1}\orcidlink{0000-0001-6478-0534}}
\authorrunning{Ruru Xu et al.}
%
\institute{Computer Engineering Department, Istanbul Technical University, Istanbul, Turkey \and
Department of Artificial Intelligence and Data Engineering, Istanbul Technical University, Turkey \and
Department of Applied Mathematics, University of Twente, Enschede, 7522 NB, The Netherlands\\
\email{xu21@itu.edu.tr}\\
\url{https://github.com/Ruru-Xu/HyperCMR}}
\maketitle              
\begin{abstract}
Accelerating image acquisition for cardiac magnetic resonance imaging (CMRI) is a critical task. CMRxRecon2024 challenge aims to set the state of the art for multi-contrast CMR reconstruction. This paper presents HyperCMR, a novel framework designed to accelerate the reconstruction of multi-contrast cardiac magnetic resonance (CMR) images. HyperCMR enhances the existing PromptMR model by incorporating advanced loss functions, notably the innovative Eagle Loss, which is specifically designed to recover missing high-frequency information in undersampled k-space. Extensive experiments conducted on the CMRxRecon2024 challenge dataset demonstrate that HyperCMR consistently outperforms the baseline across multiple evaluation metrics, achieving superior SSIM and PSNR scores. 

\end{abstract}

\section{Introduction}
\label{sec:introduction}
Multi-contrast cardiac magnetic resonance (CMR) imaging is a crucial tool for comprehensive cardiac assessment, providing detailed insights into cardiac structure and function across various contrast levels. Each contrast-weighted image—such as Cine, Aorta, Mapping, and Tagging—emphasizes different aspects of cardiac physiology, making them indispensable for accurate diagnosis. However, acquiring these images typically requires extended scan times, increasing patient discomfort and leading to more motion artifacts, complicating the reconstruction process \cite{knoll2020deep}. Reconstructing multi-contrast CMR images from undersampled k-space data is particularly challenging because it requires the preservation of intricate details specific to each contrast, despite their distinct characteristics. For example, Cine imaging captures the dynamic motion of the heart, Aorta imaging focuses on the vascular wall, and Mapping requires a precise depiction of myocardial tissue properties. These varying demands make it difficult to design a universal reconstruction model that performs well across multiple modalities. Traditional methods often compromise on essential details required for certain contrasts or struggle to generalize across different contrast types \cite{hammernik2018learning}.

To address these limitations, we introduce HyperCMR, a novel framework specifically designed to handle the complexities of multi-contrast CMR reconstruction. HyperCMR builds upon existing deep learning techniques, incorporating substantial innovations to tackle the challenges inherent in multi-contrast imaging. A key enhancement is the integration of a specialized loss function, Eagle Loss \cite{sun2024eagle}, engineered to recover the high-frequency information in undersampled k-space. 

Recent advances in deep learning have demonstrated the feasibility of reconstructing CMR images from undersampled k-space data \cite{schlemper2017deep}\cite{lustig2008compressed}\cite{xu2024segmentation}\cite{yang2024attention}. However, these methods often struggle to balance global structural coherence with preserving local details across different contrast modalities. In contrast, HyperCMR not only achieves this delicate balance but also delivers superior reconstruction performance across multiple modalities, providing a robust and effective solution for multi-contrast CMR reconstruction.

\section{Datasets}
\subsection{Data and Task Description}
The CMRxRecon2024 challenge \cite{wang2024cmrxrecon2024} provides the multi-contrast k-space data utilized in this paper. The dataset is divided into training, validation, and test sets, each containing different anatomical views and contrast settings. 

The training set includes various CMR modalities, such as Cine (196 subjects), Aorta (154 subjects), Mapping (193 subjects), and Tagging (143 subjects), each with distinct imaging characteristics that highlight different aspects of cardiac anatomy and function. Additionally, the validation and test sets introduce two previously unseen contrast modalities: Flow2d and BlackBlood, further increasing the complexity of the reconstruction task. Each modality in the validation and test sets contains data from 60 subjects.

The primary objective of this challenge is to develop a contrast-general model capable of delivering high-quality image reconstruction from highly accelerated, uniformly undersampled k-space data. HyperCMR is specifically designed to address this challenge by improving the reconstruction quality across these different modalities. The inclusion of unseen contrasts in the validation and test sets underscores the importance of generalization in model performance, which is a central focus of the HyperCMR framework.

\subsection{Datasets processing and Training idea}
\label{add:data_processing}
Based on different modalities (aorta\_sag, aorta\_tra, cine\_lax, cine\_sax, cine\_ \\
lvot, T1map, T2map, tagging) and different sizes (cine\_lax includes two sizes: [:, :, :, 204, 448] and [:, :, :,  168, 448]. cine\_sax includes three sizes: [:, :, :,  246, 512], [:, :, :,  162, 512] and [:, :, :,  204, 512]), we divide the dataset into 11 groups to facilitate us to set different batch sizes for training. The training process is shown in Algorithm \ref{alg:training_process}.

\begin{algorithm}[h]
\caption{Training process across modality groups}
\begin{algorithmic}
    \REQUIRE 11 modality groups, max\_epochs, model, optimizer
    \FOR{epoch in range(max\_epochs)}
        \STATE total\_train\_loss = 0
        \STATE total\_val\_loss = 0
        \FOR{modality\_group in [group\_1, group\_2, ..., group\_11]}
            \STATE modality\_train\_loss = train\_epoch(modality\_group, model, optimizer, ...)
            \STATE modality\_val\_loss = validate\_epoch(modality\_group, model, ...)
            \STATE total\_train\_loss += modality\_train\_loss
            \STATE total\_val\_loss += modality\_val\_loss
        \ENDFOR
        \STATE \textbf{Compute total\_train\_loss and total\_val\_loss for this epoch}
    \ENDFOR
\end{algorithmic}
\label{alg:training_process}
\end{algorithm}

\section{Methods}
\subsection{Enhancing PromptMR: The Development of HyperCMR}

\begin{figure}[t]
    \centering
    \includegraphics[width=1\linewidth]{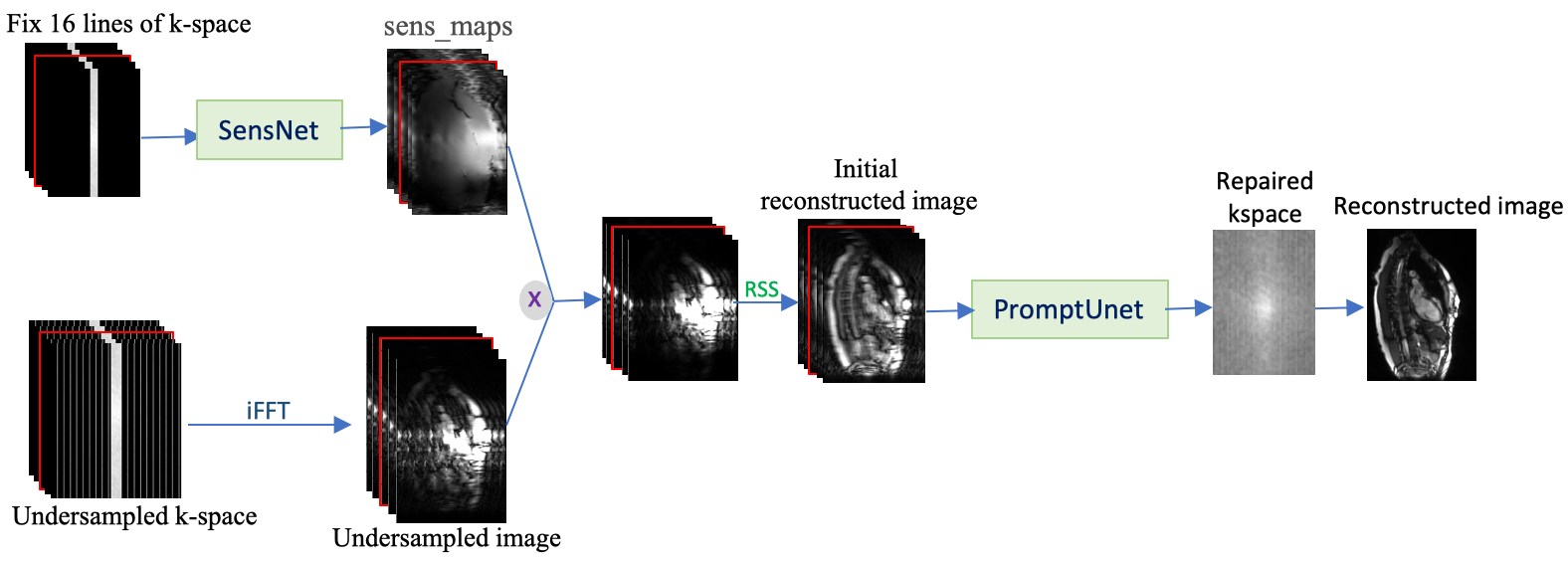}
    \caption{Overview of the HyperCMR framework. Our pipeline includes generation of sensitivity maps and repaired k-space generation with promptUnet.}
    \label{fig:method-pipline}
\end{figure}

The HyperCMR framework builds upon the established PromptMR \cite{xin2023fill} model. PromptMR is particularly effective in restoring missing information in the low-frequency regions of undersampled k-space, which contributes significantly to the overall structural integrity of the reconstructed images. However, it struggles with the recovery of high-frequency details, which are crucial for preserving the finer structures and textures in CMR images. High-frequency components typically encode critical information such as sharp edges and intricate patterns, which are vital for accurate and detailed cardiac assessment. To address this limitation, HyperCMR introduces several key enhancements, utilizing a combination of loss functions specifically designed to recover high-frequency information from undersampled k-space. By optimizing the model's ability to focus on both low- and high-frequency regions, HyperCMR achieves a more balanced and comprehensive reconstruction performance. 

As illustrated in Figure \ref{fig:method-pipline}, the HyperCMR framework leverages the temporal similarity of information by connecting slices from two adjacent time frames. With the inclusion of 10 coils, the input data format expands to 50 channels, ultimately producing a single-channel reconstructed image. The process begins by fixing 16 lines of k-space data, which are subsequently used to estimate sensitivity maps through SensNet. An initial reconstruction is performed using the inverse Fast Fourier Transform (iFFT), and the resulting image, combined with the sensitivity maps, undergoes Root Sum of Squares (RSS) calculation to yield an initial image reconstruction. This image is then refined using the PromptUnet model, which is central to both the HyperCMR and PromptMR pipelines.

\subsection{Loss Functions}

Our approach incorporates a set of advanced loss functions, each meticulously selected and weighted to balance the diverse aspects of CMR image reconstruction:

- \textbf{Data Fidelity Loss (k-space domain, \(\alpha_1 = 1.0\))}: This loss ensures that the repaired k-space data closely matches the fully-sampled ground truth, focusing on the accurate recovery of k-space information.

  \[
  L_{fidelity} = \| k_{pred} - k_{fully} \|_2^2
  \]

- \textbf{SSIM Loss (image domain, \(\alpha_2 = 1.0\))}: This component promotes structural similarity between the reconstructed image and the ground truth, ensuring perceptual quality and structural integrity.

  \[
  L_{recons\_ssim} = 1 - SSIM(img_{recons}, img_{fully})
  \]

- \textbf{Eagle Loss (image domain, \(\alpha_3 = 0.05\))}: Specifically designed to enhance high-frequency detail recovery, Eagle Loss plays a crucial role in edge preservation and the accurate reconstruction of fine structures in CMR images \cite{sun2024eagle}.

  \[
  L_{eagle} = \text{FFT-based high-pass filter loss on variance maps}
  \]

- \textbf{VGG Perceptual Loss (image domain, \(\alpha_4 = 0.1\))}: This loss guides the network to capture higher-level perceptual features, ensuring that the reconstructed image maintains visual fidelity in comparison to the ground truth \cite{simonyan2014very}.

  \[
  L_{vgg} = \sum_{l} \| VGG_l(img_{recons}) - VGG_l(img_{fully}) \|_2^2
  \]

- \textbf{Regularization Loss (k-space domain, \(\alpha_5 = 0.01\))}: This loss promotes sparsity and smoothness in the repaired k-space data, helping to prevent overfitting while preserving essential details \cite{sun2016deep}.

  \[
  L_{reg} = \| k_{recons} \|_1 + \beta \| k_{recons} \|_2
  \]

The overall loss function is a weighted combination of these components:

\[
L_{total} = \alpha_1 \cdot L_{fidelity} + \alpha_2 \cdot L_{recons\_ssim} + \alpha_3 \cdot L_{eagle} + \alpha_4 \cdot L_{vgg} + \alpha_5 \cdot L_{reg}
\]

The specific weights for each loss component were determined through extensive empirical validation. We set \(\alpha_1 = 1.0\) and \(\alpha_2 = 1.0\) to prioritize data fidelity and structural similarity, ensuring that the reconstructed image closely resembles the ground truth in both the k-space and image domains. The weight for Eagle Loss, \(\alpha_3 = 0.05\), was carefully chosen to focus on high-frequency detail recovery without overshadowing the contributions of other loss components. The VGG Perceptual Loss, weighted at \(\alpha_4 = 0.1\), is crucial for maintaining visual fidelity, particularly in capturing finer image details. Finally, the Regularization Loss, set at \(\alpha_5 = 0.01\), helps prevent overfitting while preserving key structural details, contributing to a robust and generalizable reconstruction model.

\subsection{Optimized Eagle Loss}
In our method, we improved upon the original Eagle Loss technique described in \cite{sun2024eagle} by replacing the Gaussian high-pass filter with a Butterworth high-pass filter, which preserves high-frequency information crucial for multi-contrast CMR image reconstruction. Additionally, we introduced padding and other optimizations to further enhance the performance of the loss function.

\begin{figure}[t]
\centering
\includegraphics[width=1\linewidth]{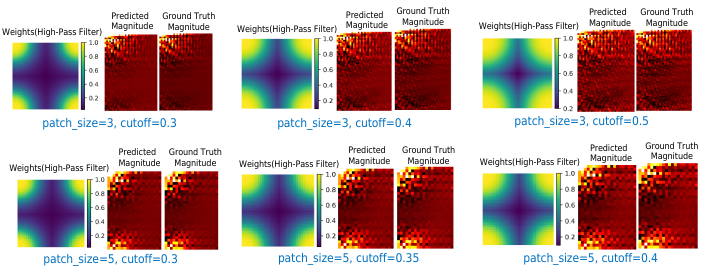}
\caption{Visualization of the impact of different patch sizes and cutoff frequencies on the high-pass filter and resulting magnitude maps. The selected patch size of 5 and cutoff of 0.35 provide balanced performance in capturing mid-level frequencies and spatial details in the HyperCMR framework.}
\label{fig:method-eagleLoss}
\end{figure}

As shown in Fig. \ref{fig:method-eagleLoss}, the effect of varying patch sizes and cutoff frequencies on the high-pass filter weights and the resulting magnitude maps within the Eagle Loss framework are as follows:
\begin{itemize}
    \item Patch Size: This parameter determines the scale of features emphasized by the high-pass filter. Smaller patch sizes (e.g., 3) focus on finer details and higher frequencies, enhancing intricate structures but potentially amplifying noise. In contrast, larger patch sizes (e.g., 5) offer a broader context, capturing mid-level frequencies while preserving essential high-frequency information, thereby maintaining structural integrity with minimal noise amplification.
    \item Cutoff Frequency: The cutoff frequency in the Butterworth filter controls the threshold for emphasizing higher frequencies. A lower cutoff frequency (e.g., 0.3) allows more high-frequency components to pass through, highlighting sharp edges and fine details, but at the cost of increased noise sensitivity. Conversely, a higher cutoff frequency (e.g., 0.4 or 0.5) suppresses very high frequencies, focusing on preserving mid-level details and reducing noise. 
    \item Selected Parameters: We selected a patch size of 5 and a cutoff frequency of 0.35 for their balanced performance. This combination effectively captures both mid-level and high-frequency details, yielding magnitude maps that closely resemble the ground truth. This balance ensures optimal structural preservation with minimal noise, which is crucial for high-fidelity reconstruction in the HyperCMR framework.
\end{itemize}

\begin{figure}
    \centering
    \includegraphics[width=1\linewidth]{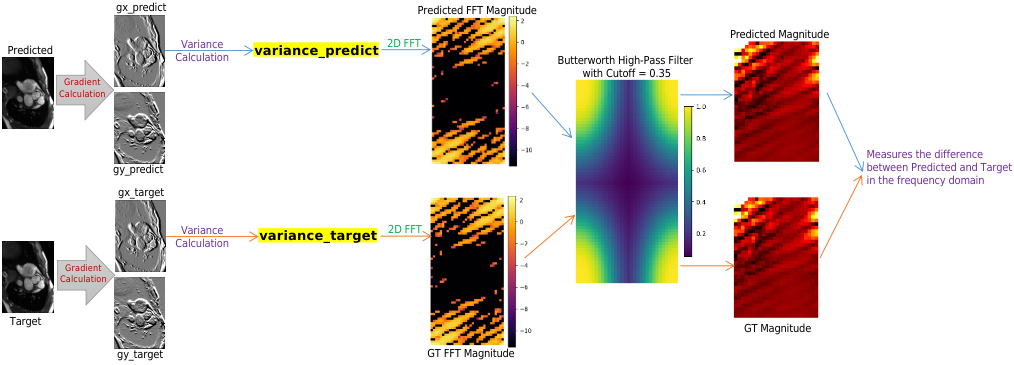}
\caption{The figure illustrates the detailed workflow of the Eagle Loss implementation for calculating gradients in the x-direction. The process begins by computing the gradients for both the predicted and target images, followed by calculating variance across non-overlapping patches. These variance maps are then transformed using a 2D FFT, and their magnitudes are filtered with a Butterworth high-pass filter. Finally, the L1 loss is calculated by comparing the filtered magnitudes in the frequency domain. The same process is applied to the y-direction gradients. The total Eagle Loss is computed as the sum of the loss for both x and y-direction gradients.}
    \label{fig:eagleLossProcess}
\end{figure}

The Eagle Loss process (Fig. \ref{fig:eagleLossProcess}), is designed to capture the structural differences between predicted and target images by focusing on their gradient information in both the x and y directions. The process begins with gradient calculation, where the predicted and target images are convolved with predefined gradient filters (Scharr kernel) to compute the horizontal (gradient\_x) and vertical (gradient\_y) gradients. These gradients highlight edge-like structures, which are critical for accurate image reconstruction.

Next, for each direction (x and y), the gradient maps are divided into non-overlapping patches, and the variance of each patch is calculated. This variance map reflects the distribution of gradient magnitudes in different regions of the image. These variance maps are then passed through a 2D Fast Fourier Transform (FFT) to obtain their frequency representations. To emphasize the high-frequency components—representing fine details—the Eagle Loss applies a Butterworth high-pass filter to the FFT magnitudes. 

The filtered FFT magnitudes of the predicted and target images are then compared using the L1 loss function in the frequency domain, capturing the differences in their high-frequency details. The overall Eagle Loss is the sum of the losses computed for both the x and y-direction gradients, ensuring that the model learns to preserve structural details across multiple directions. This approach helps the model focus on recovering fine structures and details that are often lost during undersampling in MRI reconstruction.

\section{Experiments and Results}

\subsection{Experiments}
The model was implemented in PyTorch, utilizing the AdamW optimizer with an initial learning rate of 0.00009, a weight decay of 1e-4, and a StepLR learning rate scheduler with a step size of 2 and gamma of 0.95. To manage the large computational demands, gradient accumulation was employed with 8 steps.

We trained the 4x acc, 8x acc, and 10x acc models on three servers respectively. The details are as follows:
\begin{itemize}
    \item \textbf{4x acc:} GPU: RTX 3090 24G, one epoch: 15.96 hours. The batch\_size of all 11 groups is set to 1.
    \item \textbf{8x acc:} GPU: RTX 6000 24G, one epoch: 26.13 hours. The batch\_size of all 11 groups is set to 1.
    \item \textbf{10x acc:} GPU: GV100 32G, one epoch: 20.42 hours. The batch\_size of the 11 groups (aorta\_sag, aorta\_tra, cine\_lax204, cine\_lax168, cine\_sax246, cine\_sax162, cine\_sax204, cine\_lvot, T1map, T2map, tagging) is set to 2, 2, 2, 2, 2, 1, 1, 2, 2, 4, 2 respectively.
\end{itemize}

\subsection{Results Analysis}

\begin{table}[t]
\renewcommand{\arraystretch}{1.2}
\setlength{\tabcolsep}{1.69pt} 
\centering
\small 
\caption{Performance across CMR modalities and acceleration factors (4x, 8x, 10x) in terms of SSIM, PSNR, and NMSE.}
\begin{tabular}{|c|ccc|ccc|ccc|}
\hline
\textbf{Modality} & \multicolumn{3}{c|}{\textbf{4x}} & \multicolumn{3}{c|}{\textbf{8x}} & \multicolumn{3}{c|}{\textbf{10x}} \\ \cline{2-10} 
                  & \textbf{SSIM} & \textbf{PSNR} & \textbf{NMSE} & \textbf{SSIM} & \textbf{PSNR} & \textbf{NMSE} & \textbf{SSIM} & \textbf{PSNR} & \textbf{NMSE} \\ \hline
\textbf{Tagging}  & 0.9621 & 38.57 & 0.0118 & 0.9162 & 33.98 & 0.0321 & 0.9112 & 33.20 & 0.0380 \\ \hline
\textbf{Mapping}  & 0.9508 & 37.33 & 0.0077 & 0.8880 & 31.87 & 0.0220 & 0.8791 & 30.91 & 0.0268 \\ \hline
\textbf{Cine}     & 0.9391 & 36.54 & 0.0102 & 0.8512 & 30.83 & 0.0312 & 0.8121 & 29.13 & 0.0437 \\ \hline
\textbf{Aorta}    & 0.9393 & 36.70 & 0.0137 & 0.8475 & 31.12 & 0.0452 & 0.8078 & 29.61 & 0.0609 \\ \hline
\textbf{BlackBlood} & 0.9426 & 36.14 & 0.0104 & 0.8404 & 30.22 & 0.0375 & 0.7842 & 28.30 & 0.0570 \\ \hline
\textbf{Flow2d}   & 0.9770 & 40.98 & 0.0028 & 0.9212 & 33.38 & 0.0153 & 0.9121 & 32.38 & 0.0188 \\ \hline
\textbf{Overall}  & 0.9518 & 37.71 & 0.0094 & 0.8774 & 31.90 & 0.0306 & 0.8511 & 30.59 & 0.0409 \\ \hline
\end{tabular}
\label{table:mode_result}
\end{table}

To further evaluate the robustness of our model, we analyzed its performance across specific acceleration factors, including 4x, 8x, and 10x. Table \ref{table:mode_result} provides a detailed breakdown of our model's performance across various CMR modalities under these acceleration settings. Notably, our model demonstrates consistent performance across different modalities and acceleration factors, achieving an excellent balance between maintaining structural similarity and reducing noise. 

\begin{table}[t]
\renewcommand{\arraystretch}{1.2}
\setlength{\tabcolsep}{1.69pt}
\centering
\caption{Result of 8x acceleration. Comparison of SSIM, PSNR, and NMSE scores for different loss functions. "PromptMR with SSIM Loss" represents the results obtained using the pre-trained PromptMR model, which was trained solely with SSIM loss. "Ours without Eagle Loss" refers to our model trained with a combination of multiple loss functions, excluding Eagle Loss. "Ours with Eagle Loss" indicates the results of our model trained with the same combination of multiple loss functions, including Eagle Loss.}
\begin{tabular}{|c|ccc|ccc|ccc|}
\hline
\multirow{2}{*}{\textbf{Modality}} &  \multicolumn{3}{c|}{\begin{tabular}[c]{@{}c@{}}PromptMR with \\ SSIM Loss\end{tabular}} & \multicolumn{3}{c|}{\begin{tabular}[c]{@{}c@{}}Ours without\\  Eagle Loss\end{tabular}} & \multicolumn{3}{c|}{\begin{tabular}[c]{@{}c@{}}Ours with\\  Eagle Loss\end{tabular}} \\ \cline{2-10} 
                                   & \textbf{SSIM}  & \textbf{PSNR}  & \textbf{NMSE}  & \textbf{SSIM}  & \textbf{PSNR}  & \textbf{NMSE}  & \textbf{SSIM}  & \textbf{PSNR}  & \textbf{NMSE}  \\ \hline
\textbf{Tagging}                   & 0.7423 & 28.60 & 0.1254 & 0.9155 & 33.92 & 0.0326 & 0.9166 & 34.03 & 0.0320 \\ \hline
\textbf{Mapping}                   & 0.8878 & 31.86 & 0.0223 & 0.8890 & 31.93 & 0.0218 & 0.8941 & 32.18 & 0.0210 \\ \hline
\textbf{Cine}                      & 0.8387 & 30.29 & 0.0347 & 0.8497 & 30.69 & 0.0316 & 0.8512 & 30.83 & 0.0312 \\ \hline
\textbf{Aorta}                     & 0.8311 & 30.57 & 0.0510 & 0.8457 & 31.05 & 0.0458 & 0.8487 & 31.12 & 0.0448 \\ \hline
\textbf{BlackBlood}                & 0.8394 & 30.19 & 0.0377 & 0.8422 & 30.27 & 0.0369 & 0.8426 & 30.40 & 0.0361 \\ \hline
\textbf{Flow2d}                    & 0.9078 & 32.37 & 0.0189 & 0.9205 & 33.20 & 0.0154 & 0.9212 & 33.38 & 0.0153 \\ \hline
\textbf{Overall}                   & 0.8412 & 30.65 & 0.0483 & 0.8771 & 31.84 & 0.0307 & 0.8791 & 31.99 & 0.0300 \\ \hline
\end{tabular}
\label{Table:ablation}
\end{table}

Table \ref{Table:ablation} provides a detailed analysis of the impact of different loss functions on the performance of our model at an 8x acceleration factor across various CMR modalities. Three configurations were evaluated: the results from using the pre-trained PromptMR model with SSIM loss only, our model without Eagle Loss, and our model with Eagle Loss.
\begin{itemize}
    \item Performance of Pre-trained PromptMR with SSIM Loss: The pre-trained PromptMR model, trained exclusively with SSIM loss, serves as the baseline. This model demonstrates the weakest performance across all metrics. These results indicate that while the pre-trained PromptMR model can achieve a basic level of reconstruction, it struggles with accuracy, which is crucial for precise CMR imaging.
    \item Effect of Removing Eagle Loss: Excluding Eagle Loss from our model still leads to a significant performance improvement over the baseline. This suggests that even without Eagle Loss, a combination of multiple loss functions contributes to more accurate reconstructions. However, the slightly higher NMSE values suggest that the model still encounters challenges in fully capturing high-frequency details.
    \item Impact of Including Eagle Loss: Including Eagle Loss in our model yields the best performance across all metrics and modalities. Eagle Loss effectively boosts the model’s ability to recover high-frequency details and improves overall reconstruction quality.
\end{itemize}
The ablation study demonstrates that while our model offers substantial improvements over the pre-trained PromptMR, the integration of Eagle Loss is essential for achieving state-of-the-art performance. It significantly enhances the recovery of high-frequency information in undersampled k-space, making our HyperCMR framework superior in producing high-quality, detailed reconstructions across various CMR modalities.

\begin{figure}[t]
    \centering
    \includegraphics[width=1\linewidth]{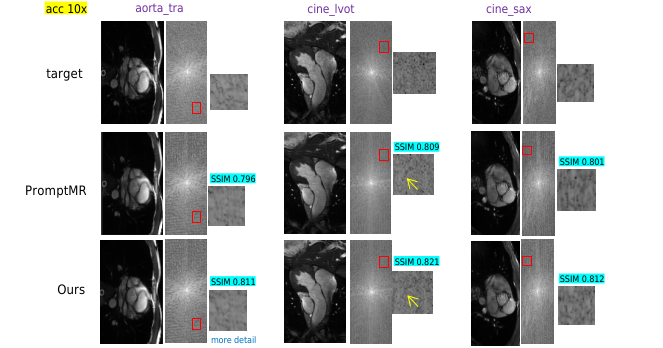}
    \caption{Visual comparison of reconstructed images for three different CMR modalities (aorta\_tra, cine\_lvot, cine\_sax) at 10x acceleration. The SSIM values and zoomed-in regions highlight the superior performance of our model (HyperCMR) compared to the PromptMR baseline. This figure provides a clear example of the improvements achieved by our method in preserving fine details and enhancing overall image quality, particularly in the high-frequency regions of the undersampled k-space.}
    \label{fig:result}
\end{figure}    

In addition to quantitative evaluations, we conducted a qualitative analysis by visually comparing the reconstructions from our proposed HyperCMR method against the baseline PromptMR model. Fig. \ref{fig:result} illustrates the reconstruction quality for three different CMR modalities: aorta\_tra, cine\_lvot, and cine\_sax, under a 10x acceleration factor. This figure demonstrates how our method effectively restores high-frequency information which is crucial for maintaining image fidelity in undersampled k-space scenarios.
In the aorta\_tra modality, the highlighted region reveals that our method preserves structural integrity more effectively than the PromptMR baseline. For the cine\_lvot modality, our method demonstrates superior preservation of k-space textures, as indicated by the yellow arrow. Additionally, in the cine\_sax modality, the superior recovery of high-frequency components further validates the efficacy of our two-stage framework in improving image fidelity and detail preservation.

\section{Conclusion}
In this paper, we introduce HyperCMR, a novel framework designed to enhance the reconstruction of highly accelerated multi-contrast cardiac magnetic resonance (CMR) images. Our approach builds on the PromptMR model and incorporates advanced loss functions, including a newly optimized Eagle Loss, specifically designed to recover the missing high-frequency information in the undersampled k-space. The success of HyperCMR can be attributed to several key innovations: the introduction of an advanced loss function framework, and the careful selection of patch size and cutoff frequency. These contributions have resulted in a powerful and generalizable model capable of effectively addressing the various challenges posed by multi-contrast CMR imaging.

Extensive experiments on the CMRxRecon2024 dataset demonstrate that our method achieves balanced performance across different CMR modalities and acceleration factors, significantly improving SSIM, PSNR, and NMSE scores. The experimental results confirm that our model is capable of preserving high-frequency information in the undersampled k-space, contributing to superior image reconstruction quality.

\begin{credits}
\subsubsection{\ackname} 

This paper has been benefitted from the 2232 International Fellowship for Outstanding Researchers Program of TUBITAK (Project No: 118C353). However, the entire responsibility of the thesis belongs to the owner. The financial support received from TUBITAK does not mean that the content of the thesis is approved in a scientific sense by TUBITAK. Computing resources used in this work were provided by the National Center for High Performance Computing of Turkey (UHeM) under grant number 4020052024. The paper also benefited from Istanbul Technical University Scientific Research Projects (ITU BAP) funds, grant numbers  FHD-2024-45302 and PMA-2024-44970.

\end{credits}
%
%
%
%

\newpage

\section*{Supplementary Material}
\appendix 
\renewcommand{\thesection}{\Alph{section}} 
In the \ref{add:task2}, we will introduce the experimental details of the Random sampling CMR reconstruction task.


\section{TASK 2: Random sampling CMR reconstruction}
\label{add:task2}
\subsection{Experimental details}
We performed distributed training on 4*A100 80G servers, randomly selecting half of datasets for training in each epoch. The batch\_size of the 11 groups (aorta\_sag, aorta\_tra, cine\_lax204, cine\_lax168, cine\_sax246, cine\_sax162, cine\_sax204, cine\_lvot, T1map, T2map, tagging) is set to 6, 6, 4, 6, 4, 4, 2, 6, 6, 10, 4 respectively. One epoch: 23.65 hours

\subsection{Results}
\begin{figure}
    \centering
    \includegraphics[width=1\linewidth]{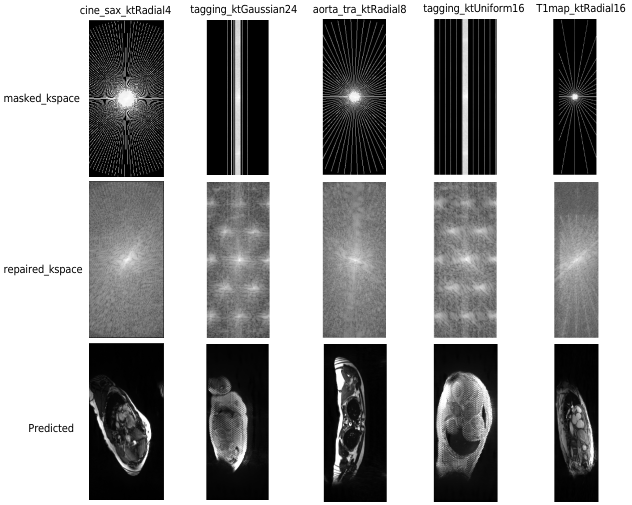}
    \caption{Task2 result. Some examples}
    \label{fig:task2Result}
\end{figure}

\end{document}